\documentclass[twocolumn,showpacs,preprintnumbers,amsmath,amssymb,nofootinbib]{revtex4}
\usepackage{graphicx}% Include figure files
\usepackage{dcolumn}% Align table columns on decimal point
\usepackage{bm}% bold math
\usepackage{mathrsfs}
\usepackage[dvips]{color}

\begin{document}

% Use /ltx/revtex4/sample/apssamp.tex as a template
%\preprint{APS/123-QED}

\title{\color[rgb]{0.2,0,0.8}
Machian strings as an alternative to dark energy}

\author{David W. Essex}
 \email{D.W.Essex@damtp.cam.ac.uk}
 \affiliation{Centre for Mathematical Sciences, Wilberforce Road, Cambridge, CB3 0WA, England}
% \homepage{http://www.homepage}
\date{\today}

\begin{abstract}
\color[rgb]{0.2,0,0.8}
The expansion history of the Universe is calculated using a simple model in which the entire rest mass energy of a massive particle is distributed throughout the set of Machian strings connecting it to all the other particles in the observable Universe. With the assumption that the energy in a Machian string has the form of positive Newtonian potential energy, the deceleration rates in the radiation and matter eras are exactly the same as in the conventional $\Lambda$CDM model. The transition from deceleration to acceleration at the present time is obtained by making the simplest possible modification of the Newtonian potential energy to represent the effect of the cosmological expansion. The effective dark matter and dark energy densities are calculated in terms of the speed of the Hubble flow at the radius of the observable Universe.

% Valid PACS numbers may be entered using the \verb+\pacs{#1}+ command.
\end{abstract}
\maketitle
\color[rgb]{0.2,0,0.8}

\section{Introduction}
It has been known for long time~\cite{brans} that the Newtonian constant of gravitation, $G$, satisfies a Machian relation of the form
\begin{eqnarray}\label{g}
\frac{GM_U}{R_Uc^2}\sim 1\,,
\end{eqnarray}
where $M_U$ and $R_U$ are, respectively, the mass and radius of the observable Universe at the present time. The purpose of the present paper is to point out that if the relation~(\ref{g}) is assumed to hold at all times then it can be used to calculate the expansion history of the Universe. 

In Section~\ref{msm}, the relation~(\ref{g}) is used to motivate a new model for an elementary particle. The calculation of the scale factor is described in Section~\ref{de} and it is shown that the entire expansion history of the Universe in the conventional $\Lambda$CDM model can be reproduced almost exactly.

\section{The Machian string model}\label{msm}
If the relation~(\ref{g}) is rewritten as 
\begin{eqnarray}\label{g2}
\frac{GmM_U}{R_U}\sim mc^2
\end{eqnarray}
then the right hand side is clearly the rest mass energy of a particle of rest mass $m$ and the left hand side may be interpreted as the magnitude of the total Newtonian gravitational interaction energy of the mass $m$ with the  observable Universe. The relation~(\ref{g2}) may be given a direct physical interpretation in terms of a new model for an elementary particle. Suppose that, instead of being point-like, an elementary particle actually consists of a point-like charged centre together with a set of Machian strings joining the centre to the centres of all other particles in the observable Universe. The entire rest mass energy of the particle is assumed to be distributed within the strings. If the strings are assumed to contain positive Newtonian potential energy, the condition that all the mass is in the strings for a particle of mass $m_i$ is 
\begin{eqnarray}\label{ams}
 \sum_j\frac{Gm_im_j}{r_{ij}}\,=\,m_ic^2\,,
\end{eqnarray}
where $r_{ij}$ is the distance between the $i^{th}$ and $j^{th}$ particles and the sum is over all particles in the observable Universe. If $\rho$ is the average mass density in the Universe then 
\begin{eqnarray}\label{mu}
 M_U\,=\,\frac{4\pi}{3}\rho R_U^3
\end{eqnarray}
and the continuum approximation gives
\begin{eqnarray}\label{gpe}
 && \sum_j \frac{m_j}{r_{ij}}\,\,\approx\,\int_0^{R_U}\!\frac{\rho}{r}\,dV\,=\,\frac{3M_U}{2R_U}\,.
\end{eqnarray}
Equation~(\ref{ams}) then becomes, for a particle of mass $m$,
\begin{eqnarray}\label{mr1}
 \frac{3GmM_U}{2R_U}\,=\, mc^2\,.
\end{eqnarray}
 
As shown below, equation~(\ref{mr1}) implies a decelerating Universe. A transition from deceleration at early times to an accelerating expansion at the present time can be obtained by including a dependence of the string energies on the rate of cosmological expansion. The simplest possible modification of the Newtonian potential energy in the string joining masses $m_i$ and mass $m_j$ has the form
\begin{eqnarray}\label{estring}
 \frac{Gm_im_j}{r_{ij}}\Big(1+a\frac{Hr_{ij}}{c}\Big)\,,
\end{eqnarray}
 where $H$ is the Hubble parameter and $a$ is a free parameter to be determined. The condition that all the mass is in the strings becomes
%\begin{eqnarray}\label{ffrac2}
 %&&\sum_j\frac{Gm_im_j}{r_{ij}}\Big(1+a\frac{Hr_{ij}}{c}\Big)\,=\,m_ic^2\nonumber\\
 %\hspace*{-0.5cm}\Rightarrow ~~~&&\frac{3GmM_U}{2R_U}\Big[1 + \frac{2a}{3}\Big(\frac{HR_U}{c}\Big)\Big]\,=\, mc^2\,.
%\end{eqnarray}
\begin{eqnarray}\label{ams2}
 \sum_j\frac{Gm_im_j}{r_{ij}}\Big(1+a\frac{Hr_{ij}}{c}\Big)\,=\,m_ic^2
\end{eqnarray}
which gives the equation
\begin{eqnarray}\label{mr2}
\frac{3GmM_U}{2R_U}\Big[1 + \frac{2a}{3}\Big(\frac{HR_U}{c}\Big)\Big]\,=\, mc^2\,.
\end{eqnarray}

\section{Calculation of the scale factor}\label{de}
The radius $R_U$ of the observable Universe is defined by the usual formula
\begin{eqnarray}\label{ru}
 R_U(t)\,=\,R(t)\int_0^t \frac{c\,ds}{R(s)}\,,
\end{eqnarray}
where $R(t)$ is the scale factor describing the expansion. The total mass $M_U$ of the observable Universe is 
\begin{eqnarray}\label{mu2}
 M_U\,=\,\frac{4\pi}{3}(\rho_b +\rho_r)R_U^3\,,
\end{eqnarray}
where $\rho_b$ is the baryon density and $\rho_r$ is the effective mass density of radiation. 
Substituting~(\ref{mu2}) into equation~(\ref{mr2}), the equation for $R(t)$ becomes 
\begin{eqnarray}\label{fsm}
2\pi G(\rho_b +\rho_r)R_U^2\Big[1 + \frac{2a}{3}\Big(\frac{HR_U}{c}\Big)\Big]\,=\,c^2\,.
\end{eqnarray}
The parameters $\Omega_b$ and $\Omega_r$ may be defined in the same way as in the standard $\Lambda$CDM model, namely
\begin{eqnarray}\label{ob}
 \Omega_b\,=\,\frac{8\pi G\rho_b^0}{3H_0^2}\hspace{0.5cm}\mbox{and}\hspace{0.5cm}
 \Omega_r\,=\,\frac{8\pi G\rho_r^0}{3H_0^2}\,,
\end{eqnarray}
 where $\rho_b^0$ and $\rho_r^0$ are the densities at the present time and $H_0$ is the present value of the
 Hubble parameter. If the scale factor is chosen so that $R=1$ at the present time then $\rho_b=\rho_b^0/R^3$ and $\rho_r=\rho_r^0/R^4$. Equation~(\ref{fsm}) then takes the form
\begin{eqnarray}\label{accel4}
 &&\hspace*{-0.5cm}\frac{3}{4}\Big(\frac{H_0R_U}{c}\Big)^{\!2}\Big(\frac{\Omega_b}{R^3}\,+\,\frac{\Omega_r}{R^4}\Big)
\Big[1 + \frac{2a}{3}\Big(\frac{HR_U}{c}\Big)\Big]\,=\,1\,.\nonumber\\
\end{eqnarray}
% where $y=R/R_0$ and $R_0$ is the value of the scale factor at the present time. 

\begin{figure}[h]
\hspace*{-3.5cm}
\includegraphics[height=6cm,width=5cm]{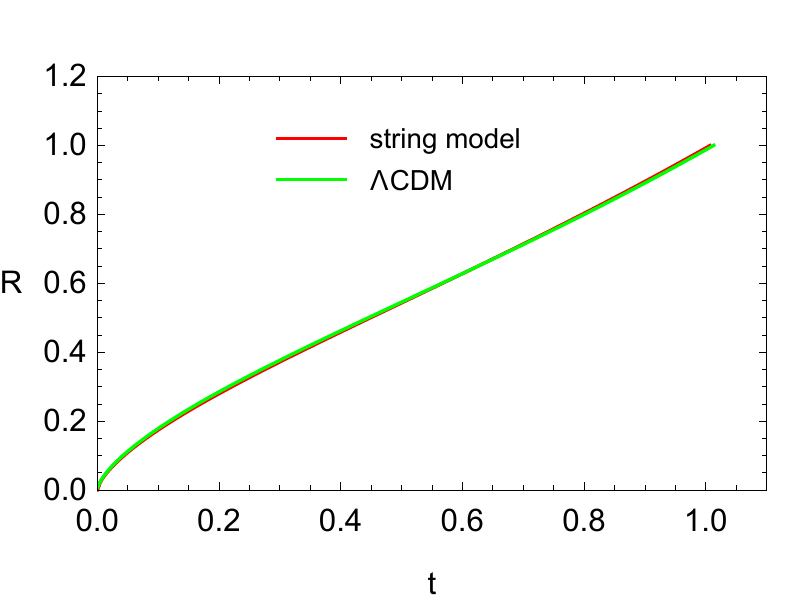}
\caption{\label{scalefactor} The scale factor $R(t)$ in the string model with $a= 0.5$ and in the $\Lambda$CDM model. The present time corresponds to $R=1$ and the two scale factors are almost identical.}
\end{figure}

The scale factor was calculated from equation~(\ref{accel4}) for a range of values of $a$ as described in Appendix~\ref{appde} using $H_0=70$\,km/s/Mpc and $\Omega_b=0.047$~\cite{bennett} and $\Omega_r=8.6\times 10^{-5}$~\cite{schneider}. For $a=0.5$, the scale factor is almost identical to the scale factor in the $\Lambda$CDM model, as shown in Figure~\ref{scalefactor}. 

\begin{figure}[h]
\hspace*{-3.5cm}
\includegraphics[height=6cm,width=5cm]{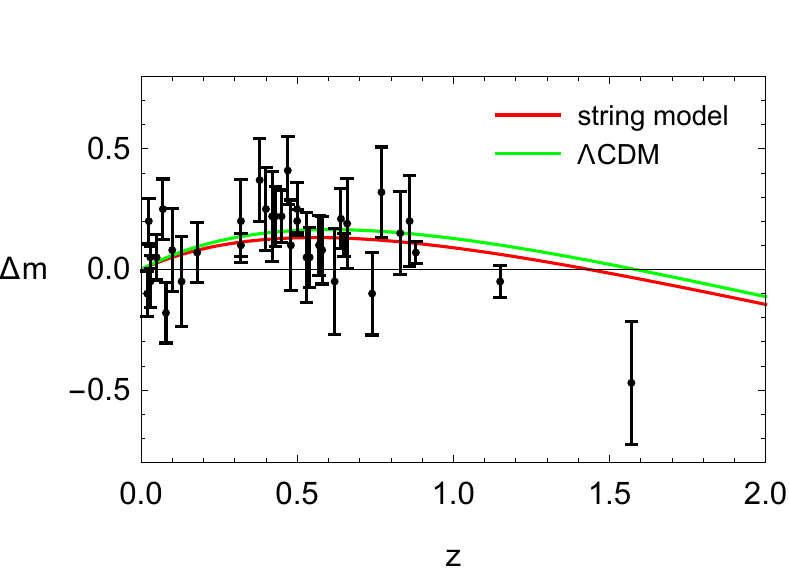}
\caption{\label{mz} Plot of residual magnitude against redshift for the string model with $a=0.5$ and the $\Lambda$CDM model, together with the experimental data points and error bars from the Supernova Cosmology Project~\cite{knop} and the High-z Supernova Search Team~\cite{riess2}.}
\end{figure}

A plot of residual magnitude against redshift for the two models is shown in Figure~\ref{mz} together with the experimental data. The acceleration parameter at the present time, $q= R\ddot R/\dot R^2$, is equal to $0.54$ in the string model with $a=0.5$ and $0.62$ in the $\Lambda$CDM model.

\begin{figure}[h]
\hspace*{-3.5cm}
\includegraphics[height=6cm,width=5cm]{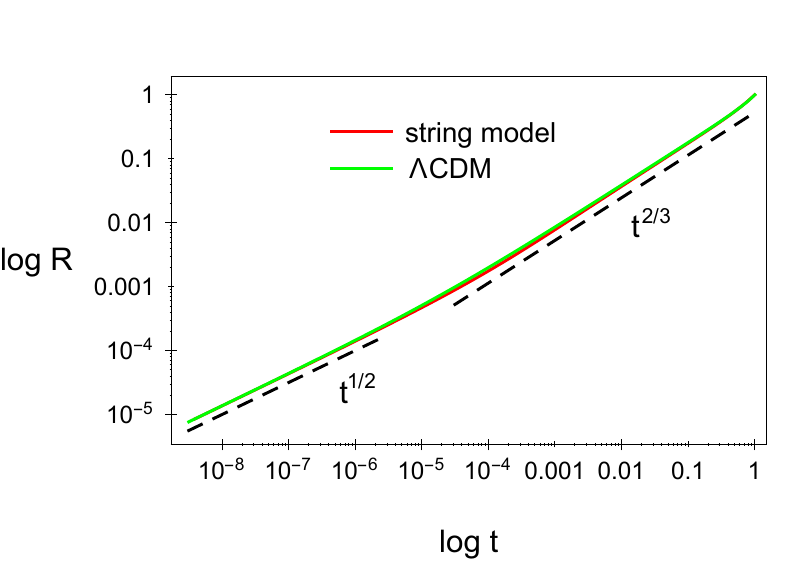}
\caption{\label{logsf} Plot of $\log R$ against $\log t$ for the best-fit scale factors $R(t)$ in the string model and in the $\Lambda$CDM model. In both models, the scale factor is proportional to $t^{1/2}$ in the radiation era and $t^{2/3}$ in the matter era.}
\end{figure}

The agreement between the string model and the $\Lambda$CDM model is even closer than suggested by Figure~\ref{scalefactor} and Figure~\ref{mz}. In addition to reproducing the acceleration of the Universe at the present time, Figure~\ref{logsf} shows that the string model also reproduces the $R\sim t^{2/3}$ time evolution of the scale factor in the matter era and the $R\sim t^{1/2}$ time evolution of the scale factor in the radiation era. 
%\section{Discussion}
%It is remarkable that the simple equation~(\ref{mr2}), with the single parameter $a$, is able to reproduce exactly the same time evolution of the scale factor in the radiation and matter eras as in the $\Lambda$CDM model and also the transition from deceleration to acceleration at late times.
%The results may be understood by referring to equation~(\ref{accel4}). At early times, $HR_U/c\ll 1$ and the factor $1+(2a/3)HR_U/c$ in~(\ref{accel4}) may be neglected. Then, to a good approximation, $R_U^2/R^4$ is constant in the radiation era and $R_U^2/R^3$ is constant in the matter era. For a power law solution $R(t)\propto t^\alpha$, equation~(\ref{ru}) gives $R_U\propto t$ for $\alpha<1$. It follows that $R(t)\sim t^{1/2}$ in the radiation era and $R(t)\sim t^{2/3}$ in the matter era, as shown in Figure~\ref{logsf}. At late times, $HR_U/c\gg 1$
%and the $\Omega_r/y^4$ radiation term in~(\ref{accel4}) is negligible so $R^3/R_U^2$ becomes proportional to $HR_U/c$, giving $HR_U^3\propto R^3$. The accelerating solution may be identified by noting that, when $R(t)$ increases faster than $t$, the integral in~(\ref{ru}) tends to a constant at late times. The ratio $R_U/R$ therefore tends to a constant and it follows from $HR_U^3\propto R^3$ that $H$ tends to a constant. The string model therefore has an accelerating solution at late times of the form $R\sim e^{Ht}$. %It is shown in Appendix~\ref{appa} that the accelerating solution is the only solution.

It is remarkable that the simple equation~(\ref{mr2}), with the single parameter $a$, is able to reproduce exactly the same time evolution of the scale factor in the radiation and matter eras as in the $\Lambda$CDM model and also the transition from deceleration to acceleration at late times.

\section{Discussion}
\subsection{The time evolution of the scale factor}
The power law solutions shown in Figure~\ref{logsf} may be derived from equation~(\ref{accel4}) as follows. Consider a decelerating solution of the form $R(t)\propto t^\alpha$, for some constant $\alpha<1$. Then $H=\alpha/t$ and equation~(\ref{ru}) gives $R_U\propto t$, from which it follows that the ratio $HR_U/c$ is constant. In the radiation era, the term proportional to $\Omega_b$ in equation~(\ref{accel4}) is negligible and, in the matter era, the term proportional to $\Omega_r$ in~(\ref{accel4}) is negligible. It follows that, to a good approximation, $R_U^2/R^4$ is constant in the radiation era and $R_U^2/R^3$ is constant in the matter era, giving $R\propto t^{1/2}$ and $R\propto t^{2/3}$, respectively.
 
The accelerating solution may be identified by noting that, when $R(t)$ increases faster than $t$, the integral in~(\ref{ru}) tends to a constant at late times. The ratio $R_U/R$ therefore tends to a constant and it follows from equation~(\ref{accel4}), since the term proportional to $\Omega_r$ is negligible, that $H$ tends to a constant. The string model therefore has an accelerating solution at late times of the form $R\sim e^{Ht}$. %It is shown in Appendix~\ref{appa} that the accelerating solution is the only solution.

\subsection{Comparison with the Friedmann equation}
In the conventional $\Lambda$CDM model, the scale factor is determined by the Friedmann equation
\begin{eqnarray}\label{fr}
H^2\,=\,\frac{8\pi G}{3}(\rho_b+\rho_r+\rho_d+\rho_\Lambda)\,,
\end{eqnarray}
where $\rho_d$ and $\rho_\Lambda$ are the contributions to the total mass density from \lq dark matter' and \lq dark energy', respectively. The Friedmann equation may be compared with equation~(\ref{fsm}), which may be written in the form 
\begin{eqnarray}\label{fsm2}
H^2\,=\,\frac{8\pi G}{3}(\rho_b +\rho_r)\frac{3}{4}\Big(\frac{HR_U}{c}\Big)^{\!2}\Big[1 + \frac{2a}{3}\Big(\frac{HR_U}{c}\Big)\Big]\,.\nonumber\\
\end{eqnarray}
Equation~(\ref{fsm2}) may be viewed as a modified Friedmann equation in which the dark matter and dark energy contributions are replaced by a dependence on the quantity $HR_U/c$, namely the speed of the cosmological expansion at the radius of the observable Universe divided by the speed of light.

In the radiation era, $\rho_r$ is much larger than all the other density components and $R(t)\propto t^{1/2}$ gives $R_U(t)=2ct$ and $H=1/2t$ so $HR_U/c=1$. Equation~(\ref{fsm2}) then gives
\begin{eqnarray}\label{fsm3}
H^2\,\approx\,\frac{8\pi G}{3}\Big(\frac{3+2a}{4}\Big)\rho_r\,,
\end{eqnarray}
which is equivalent to the Friedmann equation with a rescaled value of $\rho_r$. The Friedmann equation is recovered when $a=0.5$, which explains the value of $a$ identified in Section~\ref{de}.

In the matter era, $R(t)\propto t^{2/3}$ gives $R_U(t)\approx 3ct$ and $H=2/3t$ so $HR_U/c\approx 2$. Equation~(\ref{fsm2}) then reduces to 
\begin{eqnarray}\label{fsm4}
H^2\,\approx\,\frac{8\pi G}{3}(3+4a)\rho_b
\end{eqnarray}
which is equivalent to the Friedmann equation with a dark matter density 
\begin{eqnarray}\label{edm}
\rho_d\,=\,(2+4a)\rho_b\,.
\end{eqnarray}

In the accelerating era, the factor $HR_U/c$ becomes very large and $\rho_b$ is much larger than $\rho_r$ so equation~(\ref{fsm2}) is equivalent to the Friedmann equation with a dark energy density
\begin{eqnarray}\label{ede}
\rho_\Lambda\,=\,\frac{a}{2}\Big(\frac{HR_U}{c}\Big)^{\!3}\rho_b\,.
\end{eqnarray}
As noted above, the quantities $H$ and $R_U/R$ are constant in the accelerating era. Since $\rho_b\propto 1/R^3$, the energy density~(\ref{ede}) is constant as the Universe expands.

\section{Conclusion}
The time evolution of the scale factor in the conventional $\Lambda$CDM model can be reproduced very accurately if the entire mass-energy of every massive particle is assumed to be distributed within the Machian strings connecting it to all the other particles in the observable Universe. The deceleration in the radiation and matter eras and the transition to acceleration at the present time may all be obtained without \lq dark matter' or \lq dark energy'. Instead, there is a single parameter representing the effect of the cosmological expansion on the Newtonian potential energy in the strings.  

\section{Acknowledgements}
The hospitality of Mr Robert Buis and Mrs Joy Buis in Wartburg, South Africa, is gratefully acknowledged.

\begin{appendix}
 \section{Calculation of the scale factor}\label{appde}
\subsection{Evolution equation}
 It is convenient to define the dimensionless time, $\tau$, by $\tau= H_0t$ and the dimensionless conformal time, $\eta$, by the equation
\begin{eqnarray}\label{edef}
 \eta(t)\,=\,H_0\int_0^t \frac{ds}{R(s)}\,.
\end{eqnarray}
 Then $H_0R_U/c= R \eta$ and $HR_U/c= \dot R \eta$, where the dot denotes differentiation with respect to $\tau$.
 Substituting into~(\ref{accel4}) and solving for $\dot R$ gives
\begin{eqnarray}\label{accel5}
 \dot R \eta\,=\,\frac{2R^2}{a\Omega_b\eta^2(R+R_{eq})} - \frac{3}{2a}\,,
\end{eqnarray}
 where $R_{eq}= \Omega_r/\Omega_b$ is the value of $R$ at which the baryon and radiation densities are equal. 

\subsection{Numerical integration}
To integrate~(\ref{accel5}) numerically over several orders of magnitude in time it is convenient to change the independent variable from $\tau$ to $u= \ln\eta$, which gives
\begin{eqnarray}\label{accel6}
 \frac{dR}{du}\,=\,\frac{2R^3}{a\Omega_b\eta^2(R+R_{eq})} - \frac{3R}{2a}\,.
\end{eqnarray}
 The integration begins at the present time, $R=1$, and continues back to the singularity at $R=0$. The value of $\eta$ at the present time, $\eta=\eta_0$, is chosen so that the Hubble parameter is equal to the observed $H_0$, i.e. so that $\dot R/R= 1$. From~(\ref{accel5}), the required value of $\eta_0$ satisfies the equation
\begin{eqnarray}\label{eeta}
 \eta_0^2\Big(1\,+\,\frac{2a}{3}\eta_0\Big)\,=\,\frac{4}{3\Omega_b(1+R_{eq})}\,,
\end{eqnarray}
 which has a unique positive solution for $\eta_0$ for any non-negative value of $a$.

\begin{figure}[h]
\hspace*{-3.5cm}
\includegraphics[height=6cm,width=5cm]{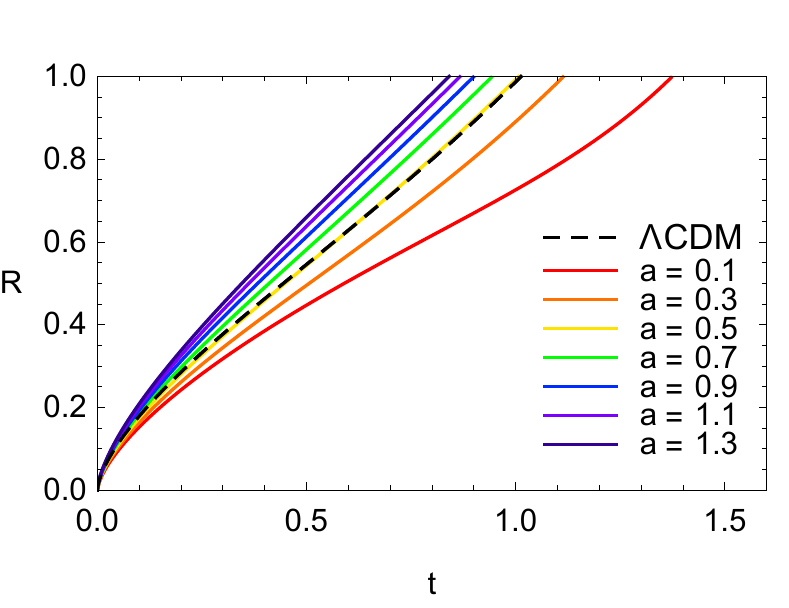}
\caption{\label{rplots} The time evolution of the scale factor in the string model, for different values of the parameter $a$, compared to the time evolution in the $\Lambda$CDM model. The present time corresponds to $R=1$. All the curves have the same slope at $R=1$ so all the models have the same value of the Hubble parameter at the present time.}
\end{figure}

\begin{figure}[h]
\hspace*{-3.5cm}
\includegraphics[height=6cm,width=5cm]{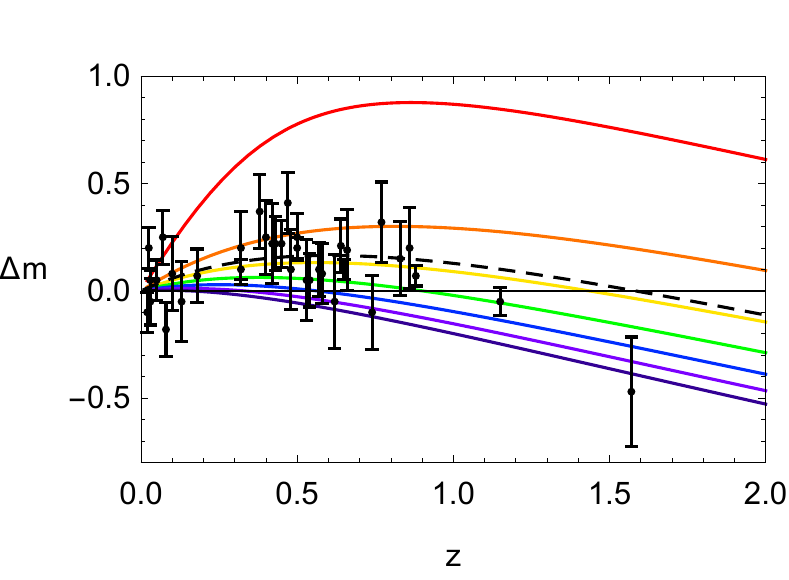}
\caption{\label{mzplots} Plots of residual magnitude against redshift for the string model, with same values of $a$ as in Figure~\ref{rplots}, and the the $\Lambda$CDM model. The data points are the same as in Figure~\ref{mz}.}
\end{figure}

The integration was performed using Mathematica~\cite{mathematica} for different values of the parameter $a$.
% \footnote{The Mathematica~\cite{mathematica} code is available upon request.}
 The resulting scale factors and the corresponding magnitude-redshift relations are shown in Figures~\ref{rplots} and~\ref{mzplots}, respectively. 
% The Mathematica~\cite{mathematica} code is available upon request.

\subsection{Values of $M_U$ and $R_U$ in the string model}
For $a=0.5$, the solution of equation~(\ref{eeta}) is $\eta_0=3.6$. Since $H_0R_U/c=R\eta$, $\eta_0$ gives the value of $HR_U/c$ at the present time. The corresponding value of $R_U$ is $15$\,Gpc, or $4.6\times 10^{26}$\,m. Neglecting the contribution from radiation, equations~(\ref{mu2}) and~(\ref{ob}) give the relation
\begin{eqnarray}\label{gmrc}
 \frac{GM_U}{R_Uc^2}\,=\,\frac{\Omega_b}{2}\Big(\frac{HR_U}{c}\Big)^{\!2}\,.
\end{eqnarray}
For $\Omega_b= 0.047$, equation~(\ref{gmrc}) gives $GM_U/R_Uc^2=0.30$, corresponding to a value of $M_U$ of $9.3\times 10^{22}$ Solar masses. The values of $HR_U/c$ and $GM_U/R_Uc^2$ are used in the accompanying paper on dark matter~\cite{paper4}.

\end{appendix}

\bibliography{master3}

\end{document}